\documentclass[aip,amsmath,amssymb,reprint]{revtex4-2}
\usepackage{graphicx}
\usepackage{dcolumn}
\usepackage{bm}

\usepackage[utf8]{inputenc}
\usepackage[T1]{fontenc}
\usepackage{mathptmx}
\usepackage{etoolbox}

\newcommand{\be}{\begin{equation}}
\newcommand{\ee}{\end{equation}}
\newcommand{\bea}{\begin{eqnarray}}
\newcommand{\eea}{\end{eqnarray}}

\begin{document}

\title{Kinetic control of competing nuclei in a dimer lattice-gas model}

\author{Dipanjan Mandal}
\email{dipanjan.mandal@warwick.ac.uk}
\affiliation{Department of Physics, University of Warwick, Coventry CV4 7AL, UK}
\author{David Quigley}
\email{d.quigley@warwick.ac.uk}
\affiliation{Department of Physics, University of Warwick, Coventry CV4 7AL, UK}
\date{\today}

\begin{abstract}
Nucleation is a key step in  the synthesis of new material from solution. Well-established lattice-gas models can be used to gain insight into the basic physics of nucleation pathways involving a single nucleus type. In many situations a solution is supersaturated with respect to more than one precipitating phase. This can generate a population of both stable and  metastable nuclei on similar timescales and hence complex nucleation pathways involving competition between the two. In this study we introduce a lattice-gas model based on two types of interacting dimer representing particles in solution. Each type of dimer nucleates to a specific space-filling structure. Our model is tuned such that stable and metastable phases nucleate on a similar timescale. Either structure may nucleate first, with probability sensitive to the relative rate at which solute is replenished from their respective reservoirs. We calculate these nucleation rates via Forward-Flux Sampling and demonstrate how the resulting data can be used to infer the nucleation outcome and pathway. Possibilities include direct nucleation of the stable phase, domination of long-lived metastable crystallites, and pathways in which the stable phase nucleates only after multiple post-critical nuclei of the metastable phase have appeared.
\end{abstract}

\maketitle

\section{\label{sec:intro}Introduction}
The phenomena of nucleation is very common in nature. Nucleation is the mechanism by which a stable phase emerges from a metastable parent phase as seen in first order phase transition. Formation of the stable phase begins by formation of sub-critical regions due to thermal fluctuations in the system. These must further grow to cross the nucleation free energy barrier before the stable phase grows to reach macroscopic size. The phenomenological description of nucleation considered most appropriate is Classical Nucleation Theory~\cite{volmer1926germ,1935-becker-aphys}. However, deviations from CNT behaviour are often observed during nucleation from solution, leading to several non-classical descriptions~\cite{erdemir2009,karthika2016review}.

Various minimal models have been used to study and gain insight into nucleation from solution~\cite{binder2016jcp}.  These include the Ising lattice-gas model~\cite{ford1999pre,cai2010pre} in the presence of external field, the Potts lattice-gas model~\cite{duff_jcp_2009,baron2014jcp,lifanov_jcp_2016,schmidt-jcp-2019,knott-jcp-2009,knott-jcp-2009}, etc. The effects of static and dynamic impurities on nucleation have also been studied in the two dimensional Ising lattice-gas~\cite{mandal2021sm}. Often, quantitative agreement between CNT (with its various modifications) and numerical simulation has been possible. Such studies can provide confidence that CNT-like consideration of the nucleation process leads to appropriate conclusions, or where such agreement is lacking, data on which to construct alternative descriptions of the nucleation process.

Very few such studies have considered situations in which multiple competing phases can emerge from the same metastable parent phase. Formation of nuclei with different structure from a multi-component solution is often observed in nature. A system studied extensively by experiment is calcium carbonate~\cite{nahi2021,darkins2022}. Under conditions where a CaCO$_{3}$ solution is supersaturated with respect to both calcite and aragonite, the relative abundance of these two polymorphs within a population of post-critical crystallites cannot be predicted purely from their relative bulk stability. A related (but different) scenario that we focus on in this work concerns a mixed solution of two solute types that can precipitate as a single-component crystal of either type. When supersaturated with respect to only one of these crystals fractional crystallisation of the most stable crystal will occur. When supersaturated with respect to both, nuclei of either crystal can form. Predicting the nucleation pathway of these competing phases is complex and requires consideration of the the respective nucleation barriers and kinetic prefactors. Anisotropy associated with the interaction and the dynamics of such particles can also play a crucial role in determining self-assembly pathways which are often non-classical~\cite{whitelam-jcp-2010}.

In this paper we report a study of this scenario via a system containing two types of interacting solute dimer.  We use on-lattice grand canonical Monte Carlo simulations under conditions where both phases are more stable than the supersaturated solution. Each type of dimer only interacts favourably with nearest neighbour dimers of same type leading to single-component space-filling structures as the precipitated phases. Our model is tuned such that both phases nucleate with very similar free energy barriers representing two phases with comparable bulk and surface free energies. We then quantify the nucleation behaviour of both phases and demonstrate how the resulting calculations can be used to infer the relative abundance of each phase in a population of post-critical precipitated crystallites. Finally we show that the population (and hence dominant phase) can be controlled by adjusting the relative rate at which dimers are transported into the system from the surrounding environmental reservoir. This indicates that solute flow constriction might be used as a control strategy where the solubility of two phases is sufficiently similar that fractional crystallisation is impractical. 

The remaining of the paper is organised as follows. In section~\ref{sec:model}, we describe the interacting dimer model and the Monte Carlo algorithm used to simulate it. Rare event sampling methods are described in section~\ref{sec:method} with computational details. Section~\ref{sec:barrier} described the process of tuning the nucleation barrier height with respect to dimer-dimer interactions. In section~\ref{sec:phase}, we study the stability of solution phase with increasing chemical potential and calculate solubility with respect to the crystalline phases. Next, in section~\ref{sec:rate}, we compute nucleation rates and estimate the relative crystallite population for different values of a parameter controlling the relative frequency at which the two dimer types are exchanged with their respective reservoirs, demonstrating control of the kinetic prefactor to nucleation. Finally we conclude in section~\ref{sec:conclusion}.

\section{\label{sec:model}Model \& Algorithm}
\begin{figure}[t!]
\includegraphics[width=\columnwidth]{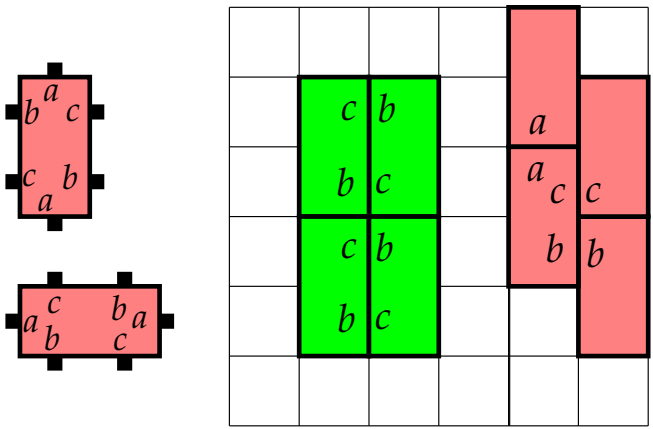}
\caption{Schematic diagram demonstrating different attachment sites $a$, $b$ and $c$ of vertical and horizontal dimers of type-A (red). Dimers of type-B (green) have similar bond structure.}
\label{fig:bond}
\end{figure}

We represent the two solute species types as dimers denoted by A (red) and B (green), on a square $L\times L$ lattice. Each dimer type occupies two adjacent squares on the lattice, in either a horizontal or vertical orientation. Dimers interact via nearest-neighbour interactions only. Each dimer has three different interaction sites (invariant under $180^{\circ}$ rotation) which we label as $a$, $b$ and $c$ as indicated in the left panel of figure Fig.~\ref{fig:bond}. Each dimer has six interactions with nearest neighbour sites. We set the energy of interaction between both dimer types and solvent (empty) lattice sites (drawn as white in Fig.~\ref{fig:bond}) to be zero, i.e. we define all other interaction energies relative to this solute-solvent interaction. 

There are six possible nearest-neighbour interactions between dimers of the same type, these being a-a, b-b, c-c, a-b, b-c and a-c. For type-A dimers we set the corresponding interaction energies $\epsilon_{aa}=\epsilon_{bb}=\epsilon_{cc}=\epsilon^A$ and the remainder are set to zero. For type-B dimers we set $\epsilon_{aa}=\epsilon_{bc}=\epsilon^B$ and the remaining interactions to zero. The list of interaction energies for type-A and type-B dimers is shown in Table \ref{table:list_energies}. For negative $\epsilon^A$ and $\epsilon^{B}$ these interactions favour formation of different 2D crystalline structures for the red (type-A) and green (type-B) dimers as shown in the right panel of Fig.~\ref{fig:bond}~\cite{nicholls2017polyomino}. We set $\epsilon^{B}=-1$ in all simulations, which defines the energy scale. All quantities with dimensions of energy (including $\epsilon^{A}$ and dimer chemical potentials) are expressed as multiples of $|\epsilon^{B}|$, i.e. the magnitude of the interaction energy between type-B sites relative to the solute-solvent interaction energy. Interaction energies between dimers of different types are also set to zero such that regions of mixed dimer types are unfavourable. 
\begin{table}
\centering
\begin{tabular}{||c c c c c c c||} 
 \hline
  & $\epsilon_{aa}$ & $\epsilon_{bb}$ & $\epsilon_{cc}$ & $\epsilon_{ab}$ & $\epsilon_{bc}$ & $\epsilon_{ca}$\\ [0.5ex] 
 \hline\hline
 type-A & $\epsilon^A$ & $\epsilon^A$ & $\epsilon^A$ & 0 & 0 & 0 \\ [0.5ex]
 \hline
 type-B & $\epsilon^B$ & 0 & 0 & 0 & $\epsilon^B$ & 0 \\ [1ex] 
 \hline
\end{tabular}
 \caption{List of different interaction energies for type-A and and type-B dimers.}
 \label{table:list_energies}
\end{table}
For brevity, we refer to the phase in which the system consists of space-filling type-A dimers as the red phase, and the phase formed from type-B dimers as the green phase.

We simulate this model in the grand canonical ensemble. The simulation grid is coupled to two separate reservoirs of type-A and type-B dimers (representing the surrounding bulk solution). These have chemical potential $\mu_A$ and $\mu_B$ respectively. We set these two chemical potentials equal in all that follows ($\mu_A=\mu_B=\mu$) to maximise competition between the two nucleating phases. 

The Monte Carlo algorithm consists of five move types, attempted with equal frequency. In the first move type we attempt to insert a dimer from the reservoir with random orientation at a random position. The dimer type is chosen randomly as type-A or type-B with equal probability. If the dimer does not overlap with an already-occupied lattice site the move is accepted with probability
\bea
P_{in}=\frac{e^{-\beta(\epsilon-\mu)}}{1+e^{-\beta(\epsilon-\mu)}}.
\eea
In the second move we attempt to remove a dimer chosen from a random site on the lattice, accepting the move with probability
\bea
P_{out}=\frac{1}{1+e^{-\beta(\epsilon-\mu)}}
\eea
if that site is occupied by a dimer. In both of the above $\epsilon$ is the binding energy of the dimer in the lattice, i.e. the energy of interaction with its nearest neighbours, and $\beta=1/k_BT$, where $T$ is the temperature and $k_B$ is the Boltzmann constant which we set to 1 throughout the simulations. Quantities such as $\beta \mu$ and $\beta \epsilon$ are hence dimensionless. 

The remaining three moves are translation, rotation and position change of a randomly selected dimer on the lattice. In the case of translation we attempt to translate a randomly chosen dimer by a number of lattice sites chosen randomly between $1$ and $L-1$ in either the horizontal or vertical direction (with equal probability of each). Rotation moves consist of rotating a randomly chosen dimer by  $90^\circ$, $180^\circ$ or $270^\circ$ keeping one end fixed. In a position change move we randomly choose a dimer from the lattice and try to deposit at another random position without change of orientation. For all three types, moves which generate overlaps with other dimers are rejected, and otherwise accepted with probability $\textrm{min}[1,e^{-\delta\epsilon/k_BT}]$, where $\delta\epsilon$ is the change in energy before and after the move. 

This move set is considered representative of a scenario in which the attachment of dimers to growing nuclei is not diffusion-limited, i.e. movement from the reservoir to the nucleus is very rapid. We will however adjust the relative microscopic dynamics of the two dimer types in what follows.

All simulations reported here are conducted with $L=100$. We refer to one simulation step as a cycle in which all five moves are attempted once on average. We define one unit of simulation time to have elapsed after $L$ moves of each type have been attempted, i.e. $5L$ moves or $L$ simulation steps.

\section{\label{sec:method}Methods}
We have performed some unbiased simulations to identify interesting parameter regimes, but are largely interested in making predictions under mild supersaturation where one would not expect to see multiple nucleation events in an $L\times L$ simulation grid within a tractable simulation time. We therefore use rare event sampling techniques to study the nucleation properties of our dimer model, specifically Umbrella Sampling (US) and Forward Flux Sampling (FFS).

We use the US~\cite{us_torrie_1977,charles-2018-us,frenkel_2004_spherical_colloids,us_review_2011} method to calculate the free energy barrier of nucleation $F(\lambda)$ as a function of cluster size $\lambda$. Dimers are considered to be part of the same cluster if they are connected via nearest neighbour interactions. The total number of connected dimers defines the cluster size $\lambda$. The quantity $F(\lambda)$ is sometimes known as droplet free energy. US is a biased sampling method to calculate the probability distribution and therefore the free energy as a function of a slow degree of freedom, in this case $\lambda$. We divide the relevant range of  ($\lambda$) into overlapping windows of equal width and calculate the free energy at each window independently. Finally we connect the free energies obtained from different windows with constant shifting to obtain the free energy barrier of the whole parameter space. The free energy of n-th window may be written as
\be
F_n^{US}(\lambda)=-k_BT\ln[P_n(\lambda)],
\ee
where $P_n(\lambda)$ is the probability of obtaining a cluster of size $\lambda$. We use an infinite square well bias potential of width equal to the width of the window. For each window we fix the width to be $20$ and the overlap with the next window is $10$. This choice of window size might not be optimal but the large overlap between two adjacent windows reduces  computational error. We calculate cluster sizes every 5 steps. The system is initialized as an otherwise empty lattice containing a nucleus of size $(w_{max}-w_{min})/2$, where $w_{max}$ and $w_{min}$ are respectively the maximum and minimum values of cluster size allowed within the window.

We calculate nucleation rates from the supersaturated solution phase to the the crystalline equilibrium phase using the FFS~\cite{2009_ffs_allen,2009_tps_escobedo,2005_ffs_allen,polymer_folding_allen_2012} method. First, we divide the range of $\lambda$ connecting liquid and crystalline phase into equally spaced interfaces. Each interface is characterised by $\lambda_i$, where $i$ is the interface index. Note that here $\lambda$ is the size of the \emph{largest} cluster unlike in US where we count all clusters of each size within the given window (n-th) to obtain $F_n^{US}(\lambda)$ as a function of  cluster size $\lambda$. The system is initialized in the metastable solution phase. We define $\lambda_S$ such that for cluster sizes less than $\lambda_S$ the system is in the metastable parent phase. We take the cluster size at the first interface $\lambda_0$ greater than $\lambda_S$. Next we calculate the initial flux at the first interface $I_0$, i.e. the number of times the first interface is crossed per unit time in an unbiased simulation. We store $N_C$ number of configurations having largest cluster size $\lambda_0$ at the first interface. Finally we calculate the interface probabilities $P(\lambda_{i+1}|\lambda_i)$ between the $i$-th interface and $(i+1)$-th as the fraction of trajectories launched from each interface that reach the next. We randomly pick a configuration previously generated at the $i$-th interface and evolve the system via our Monte Carlo dynamics until a configuration is generated with cluster size $\lambda_{i+1}$ or the largest cluster in the system becomes smaller than $\lambda_S$. The overall nucleation rate obtained from FFS can be written as
\be
I^{FFS}=I_0\prod_{i=0}^{N-1}P(\lambda_{i+1}|\lambda_i),
\label{eq:ffs_rate}
\ee
where $N+1$ is the total number of interface. Similarly, we can define the rate of generating clusters of size $\lambda$ starting from the solution phase as
\be
I(\lambda)=I_0\prod_{i=0}^{n-1}P(\lambda_{i+1}|\lambda_i),
\label{eq:ffs_rate_lambda}
\ee
where $\lambda_{n}=\lambda$ and $n\leq N$.
In our simulations, we set $\lambda_S=10$ and $\lambda_0=16$. The intermediate interfaces are set at constant gap of $10$. The total number of stored configurations at each interface is $N_C=19200$. We measure the largest cluster size $\lambda$ at every $100$ simulation steps, i.e. every time unit.

In what follows we will indicate whether $\lambda$ represents the size of red clusters, green clusters, or is agnostic to the composition of a cluster. All three cases will be employed. Detailed description of our FFS and US implementation can also be found in Ref.~\cite{mandal2021sm}.

\section{\label{sec:barrier}Tuning barrier height}
\begin{figure}[t!]
\includegraphics[width=\columnwidth]{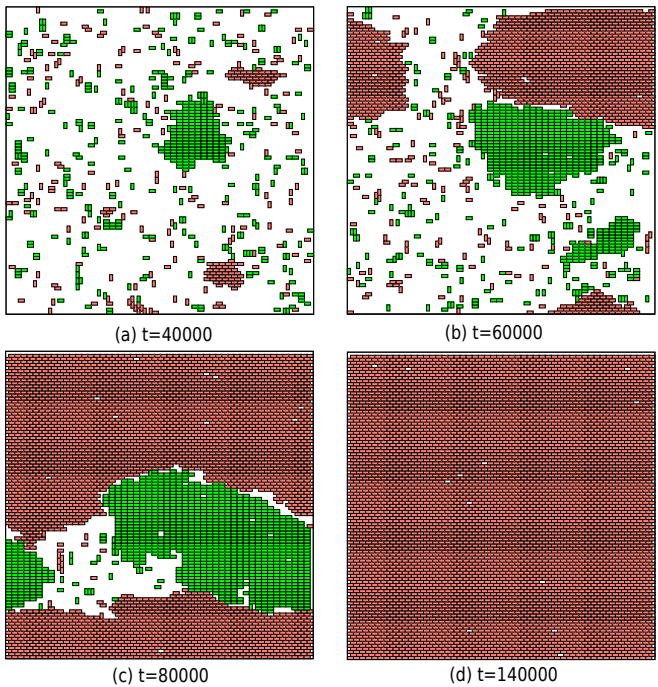}
\caption{Snapshots of typical configurations of the system of size $L=100$ at different times $t$ with $T=0.6$, $\mu=-2.86$, $\epsilon^A=-1.04$ and $\epsilon^B=-1$. At small times metastable clusters of type-B particles (green) appears which are out-competed by stable nucleus of type-A particles (red) at late times.}
\label{fig:snap_diff_t}
\end{figure}

We next seek parameters of chemical potential and dimer interaction energy which manifest closely competing nucleation between type-A (red) and type-B (green) rich crystalline phases. We begin with interactions that energetically favour formation of the red phase, and find conditions of temperature and chemical potential at which multiple nuclei can be observed in our $L=100$ simulations via unbiased method.  In Fig.~\ref{fig:snap_diff_t}(a)-(d) we have plotted snapshots of typical configurations at increasing times using parameter values $T=0.6$, $\mu=-2.86$, $\epsilon^A=-1.04$ and $\epsilon^B=-1$. The type-A rich crystalline phase which has lowest energy is the stable equilibrium phase and the type-B rich crystalline phase is metastable for these parameter values. However at small times we see the presence of type-B (green) nuclei which grow with time indicating post-critical sizes are reached [see Fig.~\ref{fig:snap_diff_t}(a)-(b) at $t=40000$ and $60000$]. At late times the stable type-A (red) phase takes over and fills the entire lattice [see Fig.~\ref{fig:snap_diff_t}(c)-(d) at $t=80000$ and $140000$]. Note that under these conditions nucleation is sufficiently rapid that growth of the metastable green phase is hindered by encountering red clusters that nucleated independently. 

\begin{figure}[t!]
\includegraphics[width=\columnwidth]{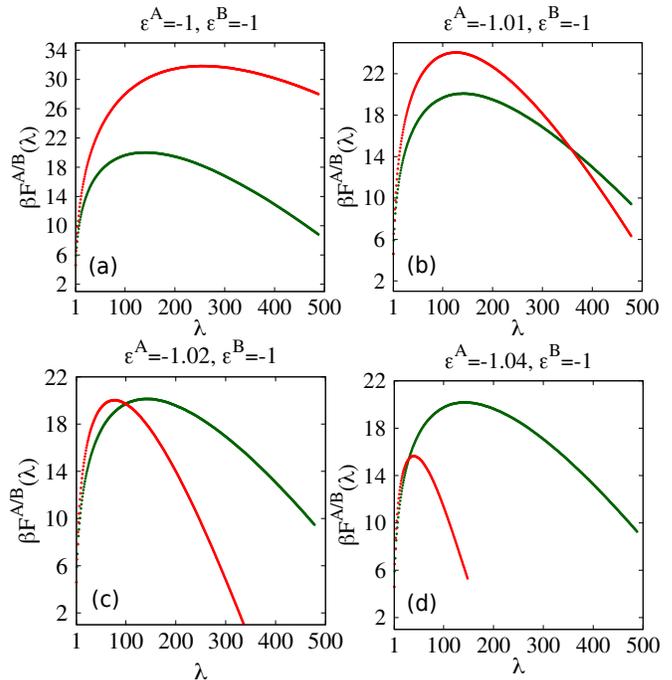}
\caption{Variation of dimensionless Free energy $\beta F^{A/B}(\lambda)$ of type-A and type-B nucleation for fixed type-B interaction energy  $\epsilon^B=-1$ and different type-A interaction energy (a) $\epsilon^A=-1$, (b) $\epsilon^A=-1.01$, (c) $\epsilon^A=-1.02$ and (d) $\epsilon^A=-1.04$ at temperature $T=0.6$ and chemical potential $\mu=-2.9$ for system size $L=100$. Red and green represent type-A and type-B respectively.}
\label{fig:free_tune}
\end{figure}
Our aim is to study conditions relevant to experiments in which a population of post-critical crystallites are observed (some stable, some metastable and long-lived). Furthermore these are spatially separated implying each has formed independently of its neighbours.  This necessitates decreasing chemical potential of the dimer reservoirs such that a substantial free energy barrier to nucleation is present in our simulation and the resulting nucleation rates are much lower than in Fig.~\ref{fig:snap_diff_t}.

We have used the US method to calculate free energy barriers to nucleation of both red and green phases. Specifically we perform two US calculations, for clusters of both type-A and type-B dimers independently. At a chemical potential of $\mu=-2.9$ barriers are sufficiently high to avoid the presence of multiple post-critical nuclei in a single simulation.

We next seek to maximise competition between the two phases. We calculate the free energy barrier for different values of $\epsilon^A$ as shown in Fig.~\ref{fig:free_tune}, again for clusters of both type-A and type-B dimers independently. Fig.~\ref{fig:free_tune}(a) corresponds to the case when the interaction energies for both type of dimers are equal $\epsilon^A=\epsilon^B=-1$. In this case the barrier height as well as the critical nucleus size are higher for type-A compared to the type-B. Here we expect to see post-critical nuclei of the type-B (green) phase emerge more rapidly than those of the type-A (red) phase because of the lower barrier height. We decrease the barrier height of type-A by decreasing the interaction parameter $\epsilon^{A}$ as shown in Fig.~\ref{fig:free_tune}(b), (c) and (d). The barrier height for nucleating both phases is comparable for $\epsilon^A=-1.02,\epsilon^B=-1$. The barrier height for nucleation of type-B becomes less than that for type-A as we decrease of interaction strength $\epsilon^A$ further to $-1.04$. In Fig.~\ref{fig:free_tune}(d) we see that $F^B(\lambda)<F^A(\lambda)$ for $\lambda<30$, i.e. there is a crossover of the two free energy curves before reaching critical size, which is absent from the other cases.
We select the case $\epsilon^A=-1.02,\epsilon^B=-1$ for all future simulations. 

\section{\label{sec:phase}Solubility}
With the selected interaction energies, the parent solution phase is clearly supersaturated with respect to both the red and green phases when coupled to reservoirs at the selected chemical potential of $\mu=-2.9$. Furthermore we expect the red (type-A rich) phase to be most stable (and have the lower solubility) as the free energy in Fig.~\ref{fig:free_tune}(c) indicates a lower free energy for red clusters at large $\lambda$.  
\begin{figure}[t!]
\includegraphics[width=\columnwidth]{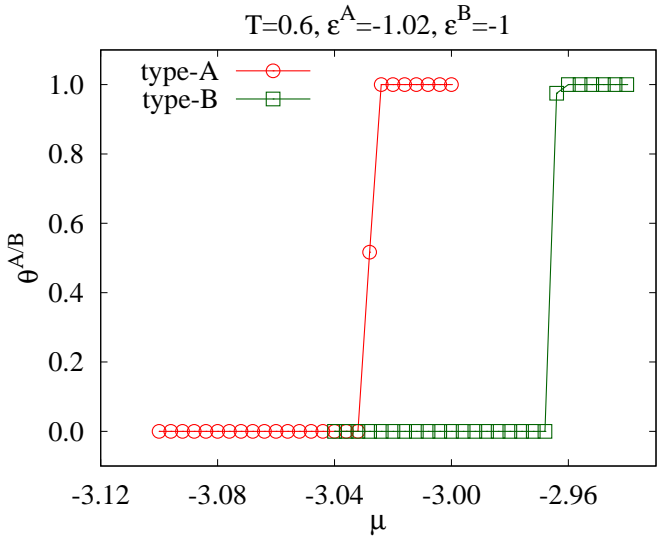}
\caption{Plots of fraction of trajectories crystallise $\theta^{A/B}$ when starting from half crystalline phase of type-A and type-B for different values of $\mu$ at $T=0.6$ with interaction energies $\epsilon^A=-1.02$ and $\epsilon^B=-1$.}
\label{fig:fraction}
\end{figure}

To quantify the supersaturation precisely, we use a two-phase coexistence method. We take a rectangular system of size $2L\times L$ with periodic boundaries. We initialise the left half of the system in one of the pure dimer-rich phases (red or green) and right half in the supersaturated solution phase. We evolve the system until the whole system transforms into either the solution phase or the dimer-rich phase of interest. We repeat the process for many realizations and measure the fraction of trajectories crystallise $\theta^{A/B}$ as a function of $\mu$. The chemical potential at which $\theta^{A/B}$ is $50\%$ defines the chemical potential at coexistence. Plots of $\theta^{A/B}$ vs $\mu$ for type-A and type-B crystalline phases are shown in Fig.~\ref{fig:fraction} confirming that the red phase is more stable.

\begin{figure}[t!]
\includegraphics[width=\columnwidth]{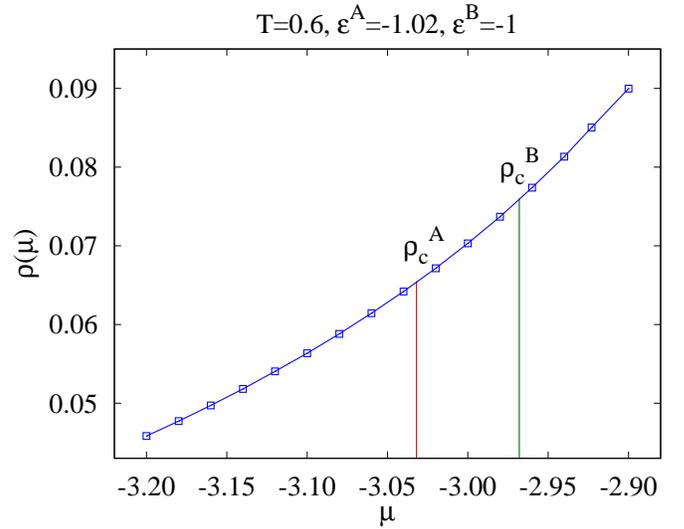}
\caption{Solution phase density $\rho$ as a function of $\mu$. For chemical potentials up to the vertical red line, both dimer-rich phases are unstable. Between the vertical red and vertical green lines only the type-A dimer-rich phase is stable.  Beyond vertical green line the type-A dimer-rich phase is stable and the type-B dimer-rich phase is metastable. At all values of $\mu$ shown the solution phase dimer density consists of type-A and type-B in equal proportions to within the accuracy of our simulations.}
\label{fig:liq_density}
\end{figure}

The supersaturation $S$ is defined as
\be
S^{A/B}(\mu)=\frac{c^{A/B}(\mu)}{c_{s}^{A/B}},
\ee
where $c^{A/B}(\mu)$ is the density (concentration) of type-A or B dimers in the supersaturated phase at chemical potential $\mu$ and $c_{s}^{A/B}$ is the density beyond which the dimers of that type are less stable in solution than in the corresponding dimer-rich phase. Density here is defined as the lattice packing fraction which can take maximum value 1. Fig.~\ref{fig:liq_density} shows the total density of dimers in the solution as a function of the reservoir chemical potential $\mu$. From this we can read off the density of the solution phase at coexistence leading to $\rho_c^A\approx 0.065$ and $\rho_c^B\approx 0.076$. As both dimers appear in equal population in the supersaturated phase for the whole range of $\mu$ plotted in Fig.~\ref{fig:liq_density}, the concentration of each dimer type $c_s^{A/B}=\rho_c^{A/B}/2$. 

Using these quantities the supersaturation at $\mu=-2.9$ (where we have tuned the two nucleation barriers to be equal) may be calculated as $S^A(-2.9)\approx1.38$ and $S^B(-2.9)\approx1.18$. 

\section{\label{sec:rate}Nucleation rates and populations}
\begin{figure}[t!]
\includegraphics[width=\columnwidth]{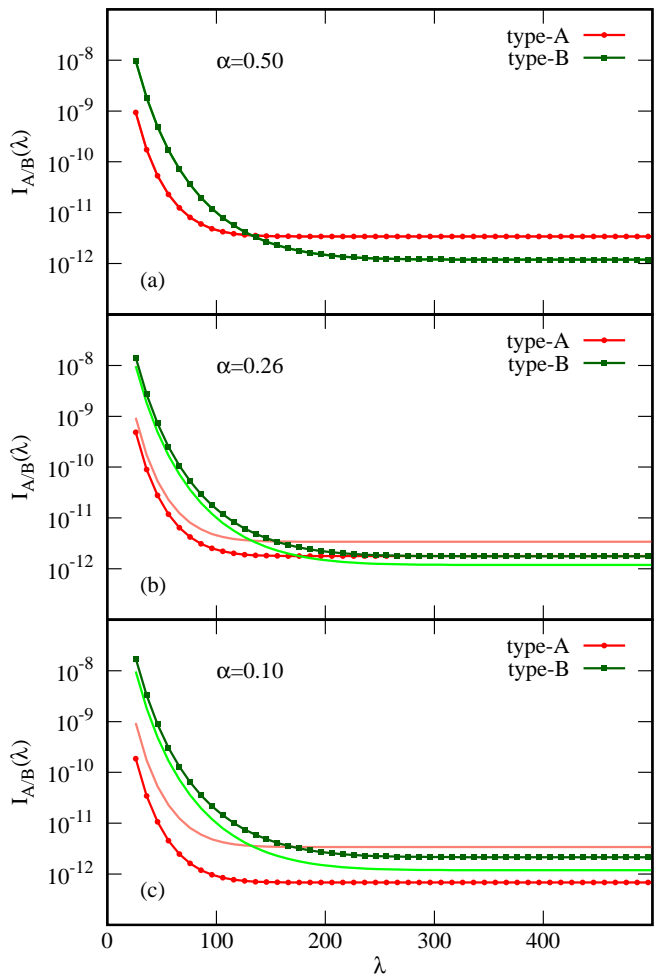}
\caption{Rate (per unit area) at which clusters of size $\lambda$ are generated in the supersaturated solution phase, $I_{A/B}(\lambda)$ for (a) $\alpha=0.50$, (b) $\alpha=0.26$ and (c) $\alpha=0.10$. For (b) and (c) the nucleation rate for $\alpha=0.5$ is plotted as a thin line without point markers for comparison. For large $\lambda$, $I_{A/B}(\lambda)$ saturates to the nucleation rate, which we denote $K_{A/B}$. One unit of area is defined as one lattice square.} 
\label{fig:rates}
\end{figure}
\begin{figure}[t!]
\includegraphics[width=\columnwidth]{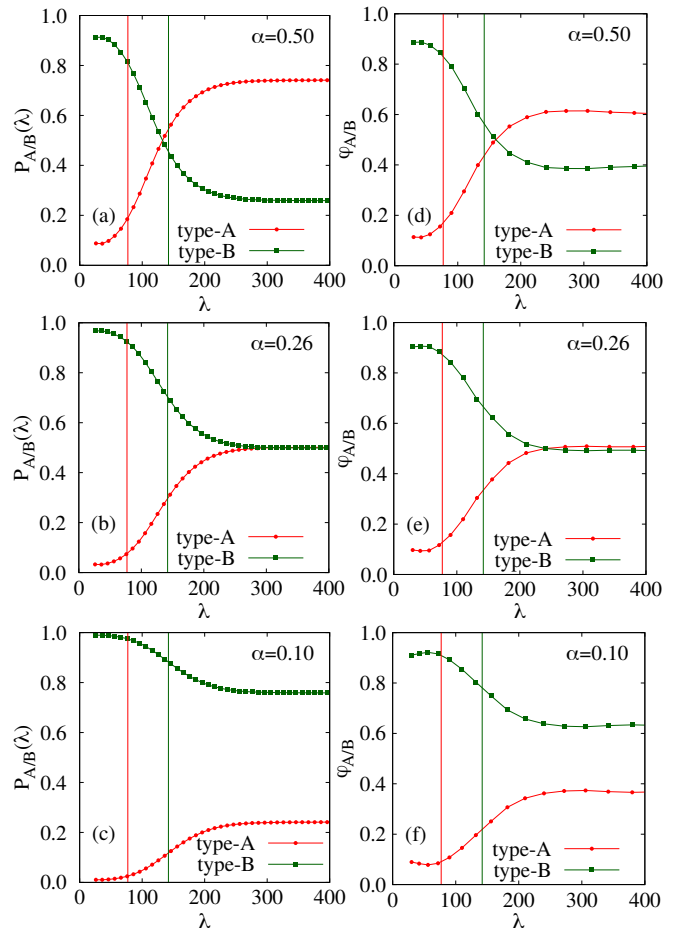}
\caption{Probability that a type-A/B cluster of size $\lambda$ is generated in the supersaturated solution before generating a type-B/A cluster of the same size $P_{A/B}(\lambda)$ for (a) $\alpha=0.50$, (b) $\alpha=0.26$ and (c) $\alpha=0.10$. Fraction of configurations within the ensemble for which the largest cluster size is $\lambda$, $\phi_{A/B}$ in which that cluster is of type-A/B in Forward Flux Sampling simulations agnostic to the solute type within each cluster. Results are plotted for (d) $\alpha=0.50$, (e) $\alpha=0.26$ and (f) $\alpha=0.10$. Vertical lines indicate the critical cluster size for each dimer-rich phase.}
\label{fig:probability}
\end{figure}
We calculate the nucleation rate for both type-A and type-B dimers via FFS for the parameter set identified in section \ref{sec:barrier} where barriers to nucleation of both type-A and type-B phases have comparable height [see Fig.~\ref{fig:free_tune}(c)]. We use a  parameter $\alpha$ to control the relative kinetics  of the two dimer types and their  frequency of exchange with their respective reservoirs. $\alpha$ can take values from 0 to 1 and determines the probability with which the five move types described in section \ref{sec:model} act on type-A vs type-B dimers. The extreme values $\alpha=0$ and $\alpha=1$ represent cases when only type-B and type-A particles are selected for trial moves. Applying this move bias to moves which preserve the number of each dimer type (i.e. canonical moves) has no measurable impact on the computed nucleation rate. When applied to exchanges with the two dimer reservoirs (grand canonical moves) the bias mimics a difference in flux of the two solute types into the depletion region around the nucleation event, representing (for example) a constriction in the flow from one of the two reservoirs. 

Fig.~\ref{fig:rates}(a)-(c) shows results for FFS calculations for three values of $\alpha$. The rates are obtained per unit time and per unit area. In each case we have performed two FFS calculations. One in which $\lambda$ is calculated as the size of the largest type-A solute cluster and other where $\lambda$ is the size of the largest type-B solute cluster. The rate of cluster generation saturates to the nucleation rate at large $\lambda$. Results indicate that it is possible to control which dimer-rich phase has the higher nucleation rate by adjusting the parameter $\alpha$. We identify $\alpha=0.26$ as the case where both dimer-rich phases have equal nucleation rate. Values of $\alpha=0.5$ and $\alpha=0.1$ represent cases in which the fastest nucleation rate is that of the type-A and type-B phase respectively. In all cases, rates at which small pre-critical clusters of type-B are generated an order of magnitude higher than the equivalent type-A rate. One can infer that whichever critical nucleus emerges first, the ensemble of pre-critical clusters from which it emerges is type-B dominated at any $\alpha$ in the range studied. Although both type-A and type-B have similar barrier heights we need to set $\alpha=0.26$ (instead of $\alpha=0.5$) to make the nucleation rates equal. This difference arises from the Zeldovic factor present in the CNT expression for nucleation rate which is proportional to the second derivative of free energy at the critical cluster size. This factor becomes smaller with increasing flatness of the free energy maximum, favouring red nucleation at $\alpha=0.5$.

For independent (i.e. spatially well-separated) nucleation events which lead to one of two phases with closely competing rates, simply computing which rate is larger will not provide sufficient information to predict the population of crystallites nucleated in each phase. To gain some insight into the population we also calculate the probability $P_{A}(\lambda)$ of generating a type-A cluster of size $\lambda$ \emph{before} any cluster of type-B reaches the same size (and vice versa). This assumes the cluster growth of type-A and B as two independent Poisson processes. The probabilities can be written as
\bea
P_A(\lambda)&=&\frac{I_A(\lambda)}{I_A(\lambda)+I_B(\lambda)},\\
P_B(\lambda)&=&\frac{I_B(\lambda)}{I_A(\lambda)+I_B(\lambda)},
\eea
where $I_A(\lambda)$ and $I_B(\lambda)$ are respectively the rate of generating a cluster of size $\lambda$ of type-A and B respectively as defined in Eq.~\ref{eq:ffs_rate_lambda}. Plots for $P_{A/B}(\lambda)$ are shown in Fig.~\ref{fig:probability}(a)-(c) for $\alpha=0.50, 0.26$ and $0.1$ respectively. We note that this assumption of independence is not always appropriate. For example, in a system where solute exchange with the environment is slow nucleation of the red phase will result in exclusion of volume accessible to green dimers in solution. This increases the effective local supersaturation of the green component, leading to correlated nucleation events. A similar enhancement of concentration has been studied in the context of gas hydrate nucleation \cite{doi:10.1021/cg401172z}. In our simulations we model rapid exchange of solute with the two dimer reservoirs and do not expect this phenomenon. However it may appear in models with more realistic solute transport. 

For $\alpha=0.1$ the picture is relatively simple. There is a $76\%$ probability that a nucleation event leading to a metastable type-B crystallite occurs before any type-A nucleation event. Type-A nucleation events that do occur will, with greater than $90\%$ probability, involve a critical type-A nucleus being formed only after a similar sized (but still pre-critical and hence transient) nucleus of type-B has already formed. Type-A nucleation events can hence only occur first by virtue of the smaller critical nucleus size. At $\alpha=0.5$ the most probable outcome of the first nucleation event is now reversed, and will be type-A with $74\%$ probability. However as at $\alpha=0.1$, critical type-A nuclei are unlikely to emerge before a type-B nucleus of the same size. By construction, at $\alpha=0.26$ the overall probability of seeing either type of nucleation event first is equally split. 

The overall picture is of a supersaturated solution in which the initial population of pre-critical nuclei is dominated by type-B, but from which either type-A or type-B nucleation may emerge first with higher probability depending on the value of $\alpha$. We stress that only some of this could be inferred from studying $F^{B}(\lambda)$ in Fig.~\ref{fig:free_tune}(c). One might argue that the lower gradient of free energy for type-B pre-critical nuclei is enough to infer that the population will favour that phase, which is correct. However the precise populations must take account of kinetic parameters also (either in a suitably modified Becker-Doring theory or via numerical calculations). To demonstrate this we have also plotted in Fig.~\ref{fig:probability}(d)-(f) the fractional populations of type-A and type-B clusters $\phi_{A/B}(\lambda)$ at each interface in FFS simulations when taking $\lambda$ to be the size of the largest solute cluster irrespective of dimer type. To generate these plots, we use a gradually increasing interface gap and slightly different success criterion at each interface. Specifically, starting from the $(i-1)^{th}$ interface, trajectories for which $\lambda_i\leq\lambda<\lambda_{i+1}$ are considered as successful. This ensures all clusters that cross (and exceed) $\lambda_i$  are counted. These populations follow, but do not match, the probabilities in Fig.~\ref{fig:probability}(a)-(c) as they do not account for the differing rate at which type-A and type-B clusters in those populations are reached.

For large $\lambda$ the rates $I_{A/B}({\lambda})$ saturate to the nucleation rates $K_{A/B}$, i.e. the rate at which critical nuclei are generated which continue to grow such that they are $100\%$ committed to the corresponding dimer-rich phase. These nucleation rates can be used to infer population statistics via the binomial distribution. Specifically the probability that a population contains $n_A$ crystallites of type-A and $n_B$ crystallites of type-B is
\be
P_{n_A,n_B}=C(n_A+n_B,n_A)\frac{{K_A}^{n_A}{K_B}^{n_B}}{[{K_A}+{K_B}]^{n_A+n_B}},
\label{eq:prob_nanb}
\ee
where $C(n_A+n_B,n_A)=\frac{(n_A+n_B)!}{n_A!n_B!}$ is the number of combinations in which $n_A$ nucleation events of type-A can appear in a sequence of $n_A+n_B$ events. For large $n_A+n_B$ this distribution is well-approximated by a normal distribution with the peak at average number of fraction of type-A crystallites
\be
\frac{n_A}{n_A+n_B}=\frac{K_A}{K_A+K_B}.
\ee 
Similarly, for type-B crystallites
\be
\frac{n_B}{n_A+n_B}=\frac{K_B}{K_A+K_B}.
\ee
This is identically the probability that the next nucleation event will be of type-A rather than type-B, i.e. the value to which the red curve saturates in Fig.~\ref{fig:probability}. For populations of independently nucleated and spatially separated crystallites this provides a route to estimate relative abundance provided there is no kinetically accessible route for transformation of post-critical nuclei of the metastable phase into the stable phase. The more in-depth data in Fig.~\ref{fig:probability} provides insight into the pre-nucleation ensemble of clusters.

In principle it should be possible to model the overall of rate at which material is formed in both type-A and type-B rich phases via an Avrami approach. Separate Avrami equations parameterised with the nucleation and growth rate of each phase could be constructed, which would be independent in the regime we study here, i.e. where nucleation events are considered spatially well separated. In other scenarios one would need to couple the two equations, such that the volume available for type-A nucleation is reduced by the presence of already-nucleated type-B material and vice versa. We defer attempts to construct such models to future work.

\section{\label{sec:conclusion}Conclusion}
We have studied the population statistics of competing nuclei in a lattice-gas model of solute precipitation in two dimensions. Solute species are represented by  two interacting dimer types. We tune the interaction energies such that the barrier height of both type-A and type-B nucleus have approximately the same value at low temperatures where nucleation is sufficiently rare that each event can be considered independent. By calculating the nucleation rates using FFS we have shown that the relative nucleation rate of two different types can be controlled by changing the relative kinetics of two types of dimers, and that the population of the two dimer rich-phases can be inferred assuming conditions of constant chemical potential. 

Generic, carefully tuned models such as this have value in investigating how various control parameters can influence the abundance of stable vs metastable crystallites in (for example) nucleation of minerals from solution. By maximising the sensitivity of selection, it should be possible to study the effect of additives (as for example in our earlier work~\cite{mandal2021sm}), nucleation at particular surfaces~\cite{2006_sear_prl}, and limitations to solute mass transport by confinement~\cite{fiona-2010,fiona-2013}. The mechanism by which these factors enhance appearance of metastable phases is often unknown. Models such as ours can potentially propose mechanisms and identify how abundance scales with these control parameters to aid experimental interpretation. Furthermore it may be possible to apply model selection techniques to infer relative nucleation rates of stable vs metastable phases from experimentally measured population statistics.

In mineral polymorphism one typically observes different crystalline structures, composed of the same atomic compositions. It would be interesting to study the nucleation of two structures from same type of particles in lattice models. 
Simple models like the model studied in this paper create mixed crystallites instead of pure structures for different choices of interaction energies when only one type of particles are present. This may be because of the model simplicity. It would be instructive to perform similar analysis for more complicated models, perhaps with different geometric shapes or more complex interactions.

\begin{acknowledgments}
We thank I. J. Ford for helpful discussions and gratefully acknowledge the use of the computational facilities provided by the University of Warwick Scientific Computing Research Technology Platform.

This research was funded in whole or in part by the EPSRC Programme Grant, EP/R018820/1. For the purpose of open access, the authors have applied a Creative Commons Attribution (CC BY) licence to any Author Accepted Manuscript version arising from this submission.
\end{acknowledgments}

\section*{Data Availability Statement}
Data generated by the simulations reported in this manuscript are available for download at \url{http://wrap.warwick.ac.uk/168078/}.

\section*{References}
%

\end{document}